\documentclass[10pt,onecolumn,tightenlines,floats,aps,amsmath,amssymb,nofootinbib,prd,floatfix,superscriptaddress]{revtex4}

\usepackage{setspace}
\usepackage{subfigure}
\usepackage{amsmath,amssymb,amsfonts,amsthm,mathrsfs}
\usepackage{graphicx,wrapfig}
\usepackage{enumerate}
\usepackage{color,xcolor}
\usepackage{hyperref}
\usepackage{physics}
\hypersetup{
	bookmarks=true,         
	unicode=false,          
	pdftoolbar=true,        
	pdfmenubar=true,        
	pdffitwindow=false,     
	pdfstartview={FitH},    
	pdftitle={Slow-roll Natural \& Hilltop Inflation in Rastall Gravity},
	pdfauthor={Ines G. Salako},
	pdfkeywords={ },
	pdfnewwindow=true,      
	colorlinks=true,        
	linkcolor=red,          
	citecolor=blue,         
	filecolor=cyan,         
	urlcolor=blue           
}

\begin{document}
	
	\title{Slow-roll Natural \& Hilltop Inflation in Rastall Gravity}
	
	\author{Isaac M. Vitohekpon}
	\email{mahutonisaacvitohekpon@gmail.com}
	\affiliation{Institut de Math\'ematiques et de Sciences Physiques (IMSP), 01 BP 613 Porto-Novo, UAC, Benin}
	
	\author{Biswajit Deb}
	\email{biswajitdeb55@gmail.com}
	\affiliation{Department of Physics, Assam University, Silchar, India}
	
	\author{Ines G. Salako}
	\email{inessalako@gmail.com}
	\affiliation{Ecole de G\'enie Rural (EGR), 01 BP 55 K\'etou, UNA, Benin}
	\affiliation{Institut de Math\'ematiques et de Sciences Physiques (IMSP), 01 BP 613 Porto-Novo, UAC, Benin}
	
	\author{V. A. Monwanou}
	\email{vincent.monwanou@imsp-uac.org}
	\affiliation{Institut de Math\'ematiques et de Sciences Physiques (IMSP), 01 BP 613 Porto-Novo, UAC, Benin}
	
	\author{Atri Deshamukhya}
	\email{atri.deshamukhya@aus.ac.in}
	\affiliation{Department of Physics, Assam University, Silchar, India}

	\begin{abstract}
	This study provides a concise analysis of inflation under Rastall gravity by examining three types of potential as the power law, natural, and hilltop potentials. Choosing a minimal interaction between matter and gravity, we derived the modified slow-roll parameters, the scalar spectral index \((n_s)\), the tensor spectral index \((n_T)\), and the tensor-to-scalar ratio \((r)\). For a general power-law potential as well as for Natural \& Hilltop inflation, we calculated these quantities and subsequently plotted their trajectories in the \((n_s, r)\) plane.  For the power-law potential, only the cases \(n = 2/3\) and \(n = 1\) satisfy the observational constraint of the Planck 2018 data. The natural potential analysis shows that the mass scale is crucial, with better compatibility achieved at \(f = 5\,M_p\) compared to \(f = 10\,M_p\). Lastly, the Hilltop potential results indicate that among the cases studied (\(m = 3/2, 2, 3,\) and \(4\)), only \(m = 3/2\) exhibits marginal consistency with observational bounds, while the other cases fail to produce acceptable \(n_s - r\) trajectories. 
		\noindent 
	
	\end{abstract}
	\maketitle

	\section{Introduction}
	\label{sec1}
	
	Observations of the cosmic microwave background radiation (CMB), both historical and recent, underline the Universe’s extensive homogeneity and flat geometry. The most intuitive and uncomplicated explanation for these features is that the Universe experienced an accelerated expansion during its earliest moments \cite{bb1}-\cite{bb4}. This inflationary epoch not only drives the rapid expansion but also creates and preserves the primordial irregularities that later evolved into the vast large-scale structures we see today. One straightforward approach to describing inflation is by incorporating a scalar field—the inflaton—into the Einstein-Hilbert action; its energy density is what propels the near-exponential expansion (e.g. \cite{bb5} and refs. therein). Recent combined data from Planck, the BICEP/Keck Array, and BAO \cite{bb6} have further narrowed the viable parameter space, showing a strong preference for concave inflationary potentials over convex ones. A simple method to express concave potentials is found in hilltop models of inflation (e.g. \cite{bb5}-\cite{bb13} and refs. therein). The major advantage of these models is that any potential with a concave profile can be effectively approximated by a hilltop potential—at least within certain parameter regions. On the flip side, the primary shortcoming is that they only offer an effective description, as hilltop potentials are intrinsically unbounded from below. A neat solution to this problem is to square the hilltop potential (e.g. \cite{bb8}-\cite{bb13}), which results in a sombrero-hat potential, similar to the approach used in symmetry-breaking inflaton models (e.g. \cite{bb5}-\cite{bb14}).

	The universe we observe originated from an initial scalar field $\vartheta > 3M_p$, where \( M_p \) represents the Planck mass in General Relativity. Yet, it is precisely the regime dominated by potential energy with an initial field value surpassing the Planck mass that is essential to accommodate the observed universe, leaving no viable alternative. The universe dominated by matter arose from cosmological perturbations at the end of the inflationary period. Moreover, recent cosmological observations indicate that the present universe is not merely expanding but is accelerating in its expansion \cite{a11,a12,a13,a14,a15,a16,a17}. A scalar field within GR falls short in explaining these observations. Hence, modifications to either the gravitational or matter sectors of Einstein’s field equations are actively sought. In turn, a new matter configuration has been proposed namely dark matter and dark energy (DE) through adjustments to the matter sector \cite{a18,a19,a20,a21,a22,a23,a24}. In pursuit of explaining the evolution of the late universe, various modified gravity theories have been examined in the literature  \cite{25}-\cite{b3}, such as \(  \mathtt{f}(R) \), \( \mathtt{f}(T) \), \( \mathtt{f}(R,T) \), \( \mathtt{f}(G) \), \( \mathtt{f}(T, \mathcal{T}) \), \( \mathtt{f}(Q) \), and others. While these theories offer descriptions of both the early inflationary period \cite{34} and the varying coupling parameters characterize the late-time epoch dominated by dark energy, a fully satisfactory gravitational theory remains undiscovered.

	In 1972, Rastall gravity \cite{38} was introduced, and now there is a burst of research aimed at integrating the recently accelerated phase of the universe in its late stage with compact astronomical entities, as documented in the literature \cite{39,40,41,42,43,44,45,46}. In this framework, the energy–momentum tensor is non-zero, a feature that Fabris et al. \cite{47} exploited to analyze the evolution of a scalar field’s gravitational potential. When considering the canonical energy–momentum tensor for a scalar field in Rastall’s theory within a cosmological setting, it is observed that, within the perturbative regime, this configuration remains consistent solely when matter is present. Furthermore, it was discovered that the ensuing coupling between the scalar field and gravity produces an equation reminiscent of those encountered in Galileon theories. Akarsu et al. \cite{48} examined an extension of the standard Cold Dark Matter (CDM) model within Rastall gravity, where two distinct modifications to the Friedmann equation are incorporated: one arising from a novel contribution of matter sources (treated as an effective source) and another due to an altered evolution of material sources. Their work also delves into how these modifications address certain low redshift tensions, including the Hubble tension associated with the standard CDM. Combining the fundamentals of scalar–tensor gravity with Rastall’s idea of violating the conventional conservation law leads to a scalar–tensor theory defined by two parameters, the Rastall parameter and the equation of state parameter ($\omega$) \cite{49}. 

The main goal of this study is to investigate how Rastall gravity impacts the cosmic inflation scenario, under the assumption of a minimal interaction between matter and gravity. Three types of potentials have been considered viz., power law potential, natural potential, and Hilltop potential in order to: i) derive the modified slow-roll parameters along with the scalar (\(n_s\)) and tensor (\(n_T\)) spectral indices, as well as the tensor-to-scalar ratio (\(r\)) ii) analyze the evolution of trajectories in the \((n_s, r)\) plane for each of these potentials iii) compare the theoretical predictions with Planck 2018 observational data. In this way, our work aims to clarify the conditions under which these inflation models, when set within the framework of Rastall gravity, can yield results that are consistent with current cosmological observations.

The paper is organized as follows: In Sec. \ref{sec2}, slow-roll inflation within General Relativity is revisited, and we introduce the slow-roll parameters, the observable spectral indices, along with other cosmological quantities of interest. Sec. \ref{sec3} details the broader equations of Ratall gravity. In Sec. \ref{sec4}, we meticulously examine the slow-roll approximation in the modified model. Utilizing the refined expressions of the slow-roll parameters, Sec. \ref{sec:Inflationary_models} applies these findings to various inflationary models. Lastly, Sec. \ref{sec:discussion} summarizes the work and presents its conclusions.
\section{Cosmological inflation scenario  in Einstein gravity }
\label{sec2}

The Einstein formulated GR action is expressed as\cite{Einstein:1916}  
\begin{equation}  
	\label{eq:actionEH}  
	S_{\textrm{GR}} = \int   \left(\frac{R}{2\kappa} + \mathcal{L}_{m}\right) \sqrt{-g} \dd[4]{x},  
\end{equation}  
By setting $\delta S_{\textrm{GR}}=0$, the equations governing the Einstein field emerge,  
\begin{equation}  
	\label{eq:EFE}  
	R_{\mu\nu} - \frac{1}{2} R g_{\mu\nu} = \kappa \Theta_{\mu\nu}  .  
\end{equation}  

The matter content is described by the tensor $\Theta_{\mu\nu}$, which is expressed as  
\begin{eqnarray}
	\label{eq:Stress-energy tensor}
	\Theta_{\mu\nu} \equiv - \frac{2}{\sqrt{-g}} \fdv{(\sqrt{-g}\mathcal{L}_m)}{g^{\mu\nu}} = g_{\mu\nu}\mathcal{L}_m - 2 \fdv{\mathcal{L}_m}{g^{\mu\nu}}.
\end{eqnarray}  

To accurately describe the present structure of our universe, the most appropriate metric is the FLRW metric, given by  
\begin{eqnarray}
	\label{eq:FLRW}
	\dd{s}^2 = -\dd{t}^2 + a(t)^2 \qty(\frac{\dd[2]{r}}{1-Kr^2} + r^2 \dd{r}^2 (\dd{\theta}^2+\sin[2](\theta)\dd{\phi}^2)),
\end{eqnarray}  
where $a(t)$ and $K$ denote, respectively, the scale factor and the spatial Gaussian curvature.  

A spatially homogeneous scalar field, known as the inflaton and represented by $\vartheta =\vartheta(t)$, can generate the simplest inflationary model. This field is governed by the Lagrangian,  
\begin{eqnarray}
	\label{eq:lagrangian_scalar_field}
	\mathcal{L}_m^{(\vartheta)} = -\frac{1}{2}\, g^{\mu\nu}\, \partial_{\mu}\, \vartheta \partial_{\nu} \vartheta - V(\vartheta) =  \frac{1}{2} \dot{\vartheta}^2 - V(\vartheta).
\end{eqnarray}  

Utilizing \eqref{eq:FLRW} in conjunction with \eqref{eq:Stress-energy tensor}, we derive:  
\begin{eqnarray}  
	\Theta_{\mu\nu}^{(\vartheta)} = \partial_\mu \vartheta \partial_\nu \vartheta + g_{\mu\nu} \qty(\frac{1}{2} \dot{\vartheta}^2 - V(\vartheta)),  
\end{eqnarray}  

where the scalar field exhibits properties analogous to a perfect fluid, characterized by the following expressions for its energy density $\rho_{\vartheta}$ and pressure $p_{\vartheta}$:  

\begin{eqnarray}  
	\label{eq:density_and_pressure_GR}  
	\Theta_{00}^{(\vartheta)}  =\frac{\dot{\vartheta}^2}{2} + V(\vartheta) = \rho_{\vartheta},\quad \Theta_{ij}^{(\vartheta)} =p_{\vartheta} g_{ij} = \qty(\frac{\dot{\vartheta}^2}{2} - V(\vartheta)) g_{ij}.  
\end{eqnarray}  

The trace can be computed as follows:  
\begin{eqnarray}  
	\label{eq:trace_T_GR}  
	\Theta^{(\vartheta)} = g^{\mu\nu}\Theta_{\mu\nu}^{(\vartheta)} = \dot{\vartheta}^2 - 4V(\vartheta).  
\end{eqnarray}

The Friedmann equation and the continuity equation can be derived as follows:  
\begin{eqnarray}  
	H^2 = \frac{\kappa \rho_\vartheta}{3}& =& \frac{\kappa}{3} \qty(\frac{\dot{\vartheta}^2}{2} + V(\vartheta)),\label{eq:1stFriedmannEq_GR}\\  
	\frac{\ddot{a}}{a} = - \frac{\kappa}{6} (3p_\vartheta+\rho_\vartheta) &=& - \frac{\kappa}{3} \qty( \dot{\vartheta}^2 - V(\vartheta)  ), \label{eq:2ndFriedmannEq_GR}\\  
	\dot{\rho}_{\vartheta} + 3H\qty(\rho_\vartheta + p_\vartheta) &=& 0 \label{eq:continuity_equation_GR}  
\end{eqnarray}  
where the Hubble parameter is defined as $H \equiv \dot{a}/a$.  

Substituting \eqref{eq:density_and_pressure_GR} into \eqref{eq:continuity_equation_GR}, we derive the Klein-Gordon equation governing the inflaton field as follows:  
\begin{eqnarray}  
	\label{eq:Klein-Gordon_GR} \ddot{\vartheta} +3\,H\,\dot{\vartheta} +
	\dv{V}{\vartheta} = 0.  
\end{eqnarray}

In the Universe’s earliest moments, its expansion is characterized by a rate parameter defined as  
$
\dv{(H^{-1})}{t} \ll 1
$
\cite{main1,main2}. In this context, the inflationary phenomenon gives rise to the following slow-roll condition \cite{main1,main2}:
\begin{eqnarray}
	\label{eq:slow--roll_condition} \dot{\vartheta}^2 \ll V(\vartheta).
\end{eqnarray}
From this condition, one can infer the primary slow-roll parameter, \(\epsilon\), which is expressed as
\begin{eqnarray}
	\label{eq:slow-roll_parameters_GR} \epsilon = -\frac{\dot{H}}{H^2} =
	3\,\dot{\vartheta}^2\,\Bigg
	(\frac{1}{\dot{\vartheta}^2\,+2\,V(\vartheta)}\Bigg),
\end{eqnarray}
with the minimal requirement for achieving inflation being \(\abs{\epsilon} \ll 1\) \cite{main1,main2}.  
Furthermore, by applying the slow-roll approximation \eqref{eq:slow--roll_condition} and employing the Friedmann equations \eqref{eq:1stFriedmannEq_GR} and \eqref{eq:2ndFriedmannEq_GR}, one can define, at first order, an analogous slow-roll parameter denoted by \(\epsilon_{\textrm{V}}\) that depends solely on the potential \(V(\vartheta)\):
\begin{eqnarray}
	\label{eq:epsilon_V_GR}
	\epsilon &\approx &\frac{3\dot{\vartheta}^2}{2V(\vartheta)} \cr
	&= &\frac{1}{2\kappa} \qty(\frac{V_{,\vartheta}}{V})^2 \cr
	&\equiv &\epsilon_{\textrm{V}}.
\end{eqnarray}

By differentiating \eqref{eq:epsilon_V_GR} with respect to time, one finds that the second slow-roll parameter, $\eta$, is given by:
\begin{eqnarray}
	\dot{\epsilon} &=& 2\frac{\dot{H}^2}{H^3} - \frac{\ddot{H}}{H^2} \cr
	&=& 2H\epsilon (\epsilon - \eta), \cr &&\eta \equiv -
	\frac{\ddot{\vartheta}}{H\dot{\vartheta}}.
\end{eqnarray}

In a similar manner to the derivation of $\epsilon_{\textrm{V}}$, the parameter $\eta_{\textrm{V}}$ can be obtained by using \eqref{eq:1stFriedmannEq_GR}, \eqref{eq:2ndFriedmannEq_GR}, and \eqref{eq:Klein-Gordon_GR}, yielding:
\begin{eqnarray}
	\label{eq:eta_V_GR} \eta_{\textrm{V}} &=& \eta + \epsilon \cr
	&\approx& \frac{1}{\kappa} \qty(\frac{1}{V}\, \dv[2]{V}{\vartheta}).
\end{eqnarray}

The spectral indices expressed in terms of the slow-roll parameters \cite{Liddle:2000,Piattela:2018} as:
\begin{eqnarray}\label{eq:spectral_indices_GR}
	n_\textrm{s}-1 &=& \dv{\ln(\Delta_{\textrm{S}}^2)}{\ln(k)} \cr
	&=& -4\epsilon + 2 \eta \cr
	&\approx& - 6\epsilon_{\textrm{V}} + 2\eta_{\textrm{V}},\cr
	n_\textrm{T} &=& \dv{\ln(\Delta_{\textrm{T}}^2)}{\ln(k)} \cr
	&=& -2 \epsilon \cr
	&\approx & -2\epsilon_{\textrm{V}},\cr
	r_{*} &=&
	\frac{\Delta_{\textrm{T}}^2(k_{*})}{\Delta_{\textrm{S}}^2(k_{*})}
	\cr &=& 16\epsilon \cr &\approx& 16 \epsilon_{\textrm{V}}.
\end{eqnarray}

Under the slow-roll approximation, the number of e-folds, $N$, is given by:
\begin{eqnarray}
	\label{eq:e-folds_GR}
	N &=& \int_{t_1}^{t_2} H\dd{t} \cr
	&=& \int_{\vartheta_{\textrm{f}}}^{\vartheta} \frac{H}{\dot{\vartheta}} \dd{\vartheta} \cr
	&\approx& \kappa \int_{\vartheta_{\textrm{f}}}^{\vartheta} \frac{V(\vartheta')}{V_{\vartheta}(\vartheta')} \dd{\vartheta'},
\end{eqnarray}
where $\vartheta_{\textrm{f}}$ represents the value of the inflaton at the end of inflation.

\section{Einstein Rastall Formalism}\label{sec3}

As noted in the Introduction, Rastall proposed an alternative approach to non conservative gravity by suggesting that the gravitational equations can be derived from a modification of the conservation laws \cite{Rastall1973,Rastall1976}. Initially, Rastall's fundamental equations were expressed as:
\begin{eqnarray}
	\label{re1}
	R_{\mu\nu} - \frac{\lambda_{Ras}}{2}g_{\mu\nu}R &=& \kappa \Theta_{\mu\nu},\\
	\label{re2} {\Theta^{\mu\nu}}_{;\mu} &=& \frac{1-\lambda_{Ras}}{2\kappa}R^{;\nu},\label{eqnonconserv}
\end{eqnarray}
where $\lambda_{Ras}$ is the Rastall parameter. Moreover, $\kappa$ represents the gravitational coupling constant in Rastall's theory, which is distinct from Einstein's gravitational coupling constant \cite{Rastall1973,al1,al2}. In fact, one has $\kappa=\frac{\kappa_E}{\alpha}$, where $\kappa_E=8\pi G$ is the Einstein coupling constant, and $\alpha$ is an unknown constant to be determined by other physical considerations and observations \cite{al1,al2}. It is worth noting that when $\lambda_{Ras} = 1$ (corresponding also to $\alpha=1$), the standard general relativity is recovered with the usual energy-momentum conservation. These equations can be equivalently rewritten as
\begin{eqnarray}
	\label{re1bis}
	R_{\mu\nu} - \frac{1}{2}g_{\mu\nu}R &=& \kappa\left\{\Theta_{\mu\nu} - \frac{\gamma-1}{2}g_{\mu\nu}\Theta\right\},\\
	\label{re2bis} {\Theta^{\mu\nu}}_{;\mu} &=& \frac{\gamma-1}{2}\Theta^{;\nu},
\end{eqnarray}
with
\begin{equation}
	\gamma = \frac{3\lambda_{Ras} - 2}{2\lambda_{Ras} - 1}.
\end{equation}
Here again, $\gamma = 1$ (i.e., $\lambda_{Ras} = 1$) yields General Relativity. Alternatively, one may cast these relations in the following form:
\begin{eqnarray}
	\label{e1}
	R_{\mu\nu} &=& \kappa\left\{\Theta_{\mu\nu} - \frac{2-\gamma}{2}g_{\mu\nu}\Theta\right\},\\
	\label{e2} {\Theta^{\mu\nu}}_{;\mu} &=& \frac{\gamma-1}{2}\Theta^{;\nu},
\end{eqnarray}
where $\lambda_{Ras}$ still appears as Rastall's parameter and setting $\lambda_{Ras}=1$ recovers GR. It is also feasible to derive the above equations from a Lagrangian formulation \cite{Smalley1984}. For a radiative fluid, where $\Theta=0$ and consequently $R=0$, the cosmological evolution during the radiative era mirrors that of the standard scenario. Similarly, a single fluid inflation model driven by a cosmological constant behaves as in GR. Hence, Rastall cosmologies represent an intriguing deviation from the standard cosmological model starting from the matter-dominated phase \cite{Batista:2010nq,Fabris:2011rm,Fabris:2011wz,Capone:2009xm}.

The functions $\nu$ and $\lambda$ satisfy the Rastall field equations,
\begin{equation}
	G_{\mu\nu}= T_{\mu\nu}^{eff},\label{field}
\end{equation}
with
\begin{equation}
	T_{\mu\nu}^{eff}=\kappa\left\{\Theta_{\mu\nu} - \frac{\gamma-1}{2}g_{\mu\nu}\Theta\right\}.\label{eq:Effective_T_munu}
\end{equation}
In this context, we adopt units such that $G=\alpha$, which implies $\kappa = 8\pi$ \cite{al1,al2}.

\section{Einstein--Rastall gravity within the Slow--Roll inflationary framework}
\label{sec4}

At this stage, we proceed to evaluate the components of the effective energy-momentum tensor as formulated in \eqref{eq:Effective_T_munu}, given by:
\begin{align}
	\label{eq:Effective_rho} T_{00}^{(\textrm{eff})} &=
	\frac{\kappa}{2\lambda_{Ras}-1}\left[\frac{3\lambda_{Ras}-2}{2}\,\dot{\vartheta}^2+V(\vartheta)\right]
	\equiv \rho_{\vartheta}^{(\textrm{eff})}, \\
	T_{ij}^{(\textrm{eff})} &=   \frac{\kappa}{2\lambda_{Ras}-1}\,g_{ij}\left[\frac{\lambda_{Ras}}{2}\,\dot{\vartheta}^2-V(\vartheta)\right]
	\equiv p_{\vartheta}^{(\textrm{eff})} g_{ij},
	\label{eq:Effective_pressure}
\end{align}
where $T^{(\textrm{eff})}_{\mu\nu} = 0$ holds for $\mu\neq \nu$. From this, we extract:
\begin{eqnarray}
	\label{eq:f(R,T)_EoS} w^{(\textrm{eff})} = \frac{p_{\vartheta}^{(\mathrm{eff})}}{\rho_{\vartheta}^{(\mathrm{eff})}}
	=\frac{\frac{\lambda_{Ras}}{2}\,\dot{\vartheta}^2-V(\vartheta)}
	{\frac{3\lambda_{Ras}-2}{2}\,\dot{\vartheta}^2+V(\vartheta)}.
\end{eqnarray}
Additionally, the trace of the effective energy-momentum tensor is computed as:
\begin{eqnarray}
	\label{eq:Effective_T_munu_Trace} T^{(\textrm{eff})} &=&
	g^{\mu\nu}T^{(\textrm{eff})}_{\mu\nu} \cr &=& \frac{\kappa}{2\lambda_{Ras}-1}\left[\dot{\vartheta}^2-4V(\vartheta)\right] \cr &=&
	3p_{\vartheta}^{(\textrm{eff})} - \rho_{\vartheta}^{(\textrm{eff})} \cr
	&=& (3w^{(\textrm{eff})}-1)\rho_{\vartheta}^{(\textrm{eff})}.
\end{eqnarray}

The modified Friedmann equations take the form:
\begin{align}
	H^2 &= \frac{\kappa}{3} \rho_{\vartheta}^{(\textrm{eff})} = \frac{\kappa}{3(2\lambda_{Ras}-1)}\left[\frac{3\lambda_{Ras}-2}{2}\,\dot{\vartheta}^2+V(\vartheta)\right], \label{eq:f(R,T)_Friedmann_1}\\
	\frac{\ddot{a}}{a} &= - \frac{\kappa}{6}
	(3p_{\vartheta}^{(\textrm{eff})} + \rho_{\vartheta}^{(\textrm{eff})} ) =-\frac{\kappa}{6(2\lambda_{Ras}-1)}\Bigl[(3\lambda_{Ras}-1)\dot{\vartheta}^2-2V(\vartheta)\Bigr],   \label{eq:f(R,T)_Friedmann_2}\\
	\dot{H} &= \frac{\ddot{a}}{a} - H^2 = -\frac{\kappa}{2}
	(p_{\vartheta}^{(\textrm{eff})}  + \rho_{\vartheta}^{(\textrm{eff})} ).\label{eq:f(R,T)_H_dot}
\end{align}

By referring to these relations, the continuity equation emerges along with \eqref{eq:f(R,T)_H_dot}:
\begin{eqnarray}
	\label{eq:f(R,T)_Klein-Gordon} \dot{\rho}_{\vartheta}^{(\textrm{eff})}
	+ 3H\qty(\rho_\vartheta^{(\textrm{eff})} +
	p_{\vartheta}^{(\textrm{eff})}) &=& 0\cr
	(3\lambda_{Ras}-2)\,\ddot{\vartheta}+3H\,(2\lambda_{Ras}-1)\,\dot{\vartheta}+V'(\vartheta)&=&0.
\end{eqnarray}

By implementing the slow-roll assumptions, we arrive at the first slow-roll parameter under Rastall gravity:
\begin{eqnarray}
	\label{eq:f(R,T)_epsilon} \tilde{\epsilon} = -\frac{\dot{H}}{H^2}\approx\frac{3(2\lambda_{Ras}-1)\,\dot{\vartheta}^2}{2V(\vartheta)}.
\end{eqnarray}
Furthermore, the first potential slow-roll parameter, corrected due to the Rastall contribution, follows as:
\begin{eqnarray}
	\label{epsilonv} \tilde{\epsilon} \approx 
	\frac{3(2\lambda_{Ras}-1)\,\dot{\vartheta}^2}{2V(\vartheta)}
	\simeq \frac{3(2\lambda_{Ras}-1)}{2V(\vartheta)}\;\frac{(2\lambda_{Ras}-1)\,V_{,\vartheta}^2}{3\kappa\,V(\vartheta)}
	=\frac{(2\lambda_{Ras}-1)^2}{2\kappa}\,\left(\frac{V_{,\vartheta}}{V(\vartheta)}\right)^2 \equiv \tilde{\epsilon}_{\textrm{V}}.
\end{eqnarray}

Similarly, we extract the second slow-roll parameter:

\begin{eqnarray}
	\tilde{\eta} = -\frac{1}{H}\frac{\ddot{\vartheta}}{\dot{\vartheta}} \simeq \frac{(2\lambda_{Ras}-1)}{\kappa\,V(\vartheta)}\,V_{,\vartheta\vartheta} - \frac{(2\lambda_{Ras}-1)^2}{2\kappa}\left(\frac{V_{,\vartheta}}{V(\vartheta)}\right)^2.
\end{eqnarray}

Similarly, the corrected version $\tilde{\eta}_V$ due to Rastall's influence is given by:

\begin{eqnarray}
	\label{etav} \tilde{\eta}_{\textrm{V}} &\equiv& \tilde{\epsilon} +
	\tilde{\eta} \cr &\approx& \frac{(2\lambda_{Ras}-1)}{\kappa\,V(\vartheta)}\,V_{,\vartheta\vartheta}.
\end{eqnarray}

For inflation to result in an isotropic and homogeneous universe, the number of e-folds, $\tilde{N}$, is determined from \eqref{eq:Effective_rho} and \eqref{eq:Effective_pressure} as:
\begin{eqnarray}
	\label{eq:f(R,T)_e-folds_definition} \tilde{N} &=& \int
	\frac{H}{\dot{\vartheta}} \dd{\vartheta} \cr &\approx& \frac{\kappa}{(2\lambda_{Ras}-1)} \int_{\vartheta_{\textrm{f}}}^{\vartheta{i}}
	\frac{V(\vartheta)}{V_{,\vartheta}}\,\dd{\vartheta}.
\end{eqnarray}
Here, $\vartheta_{f}$ is evaluated under the assumption that
$\bar{\epsilon}_v=1$. Finally, the expressions for the CMB parameters are obtained as follows:
\begin{eqnarray}
	\label{spectralindex}
	n_s-1&=&-6 \bar{\epsilon}_v+2\bar{\eta}_v, \cr
	r&=&16\bar{\epsilon}_v, \cr
	n_t &=&-2\bar{\epsilon}_v.
\end{eqnarray}

\section{Contrasting Rastall gravity inflationary scenarios}
\label{sec:Inflationary_models}

\subsection{Power Law Potentials}

As an initial illustration of an inflationary scenario, we consider one of the simplest cases by employing a power law potential of the form
\begin{equation}
	\label{eq:power_law_potential}
	V(\vartheta) = \lambda \vartheta^n,
\end{equation}
where \(\lambda\) is a coupling constant. Using the definitions of \(\epsilon_{\textrm{V}}\) and \(\eta_{\textrm{V}}\) provided in equations \eqref{eq:epsilon_V_GR} and \eqref{eq:eta_V_GR}, we find the slow-roll parameters to be
\begin{equation}
	\epsilon_{\textrm{V}} = \frac{n^2}{2\kappa \vartheta^2}, \quad \eta_{\textrm{V}} = \frac{n(n-1)}{\kappa \vartheta^2}\,.
\end{equation}
Since inflation terminates when \(\epsilon_{\textrm{V}}(\vartheta_{\textrm{end}})=1\), it follows that \(\vartheta_{\textrm{end}} = \frac{n}{\sqrt{2\kappa}}\). Consequently, the number of e--folds is determined from equation \eqref{eq:e-folds_GR} by
\begin{equation}
	N = \frac{\kappa}{2n}\Bigl(\vartheta^2 - \frac{n^2}{2\kappa}\Bigr)\,.
\end{equation}
Thus, one may express the slow-roll parameters in terms of the number of e-folds as
\begin{equation}
	\epsilon_{\textrm{V}} = \frac{n}{4N+n}, \quad \eta_{\textrm{V}} = \frac{2(n-1)}{4N+n}\,.
\end{equation}
Furthermore, the spectral indices for this model are derived from \eqref{eq:spectral_indices_GR}:
\begin{subequations}
	\label{eq:spectral_indexes_power_law_GR}
	\begin{align}
		n_{\textrm{S}} - 1 &\approx 2\eta_{\textrm{V}} - 6\epsilon_{\textrm{V}} = - \frac{2(n+2)}{4N+n}\,,\\[1mm]
		n_{\textrm{T}} &\approx -2\epsilon_{\textrm{V}} = -\frac{2n}{4N+n}\,,\\[1mm]
		r &\approx 16\epsilon_{\textrm{V}} = \frac{16n}{4N+n}\,.
	\end{align}
\end{subequations}

Next, we evaluate how Rastall Gravity modifies the slow-roll parameters. From equations \eqref{epsilonv} and \eqref{etav}, we obtain

	\begin{align}
	\tilde{\epsilon}_{\textrm{V}} &= \frac{(2\lambda_{Ras} - 1)^2 n^2}{2\kappa \vartheta^2}, \\
	\tilde{\eta}_{\textrm{V}} &= \frac{(2\lambda_{Ras} - 1) n(n-1)}{\kappa \vartheta^2}.
\end{align}
and 
\begin{align}
	n_s - 1 &= -\frac{6(2\lambda_{Ras} - 1)^2 n^2}{2\kappa \vartheta^2} + \frac{2(2\lambda_{Ras} - 1) n(n-1)}{\kappa \vartheta^2}, \\
	r &= \frac{16(2\lambda_{Ras} - 1)^2 n^2}{2\kappa \vartheta^2}, \\
	n_t &= -\frac{2(2\lambda_{Ras} - 1)^2 n^2}{2\kappa \vartheta^2}.
\end{align}

Here, inflation ends when the modified slow-roll parameter \(\tilde{\epsilon}_{\textrm{V}} \approx 1\), implying that
\begin{equation}
\vartheta_{\textrm{f}} = \frac{(2\lambda_{Ras} - 1) n}{\sqrt{2\kappa}}\,.
\end{equation}
Thus, according to the definition in \eqref{eq:f(R,T)_e-folds_definition}, the number of e--folds \(\tilde{N}\) is given by
\begin{eqnarray}
	\tilde{N} &=& \frac{\kappa}{(2\lambda_{Ras}-1)} \int_{\vartheta_{\textrm{f}}}^{\vartheta{i}}
	\frac{V(\vartheta)}{V_{,\vartheta}}\,\dd{\vartheta} \cr
	&=& \frac{\kappa \vartheta^2}{2n(2\lambda_{Ras} - 1)} - \frac{(2\lambda_{Ras} - 1) n}{4} \,.
\end{eqnarray}
In this context, the corrected slow-roll parameters can be rewritten as
\begin{subequations}
	\begin{align}
	\tilde{\epsilon}_{\textrm{V}} &= \frac{(2\lambda_{Ras} - 1) n}{4\tilde{N} + (2\lambda_{Ras} - 1)n}, \\
	\tilde{\eta}_{\textrm{V}} &= \frac{2(n-1)}{4\tilde{N} + (2\lambda_{Ras} - 1)n}.
\end{align}
\end{subequations}
Thus, the spectral indices in Rastall gravity are obtained from equation \eqref{spectralindex} as
\begin{subequations}
	\label{eq:spectral_indices_f(R,T)_power_law}
	\begin{align}
		n_s - 1 &= \frac{-6(2\lambda_{Ras} - 1)n + 4(n-1)}{4\tilde{N} + (2\lambda_{Ras} - 1)n}, \\
		r &= \frac{16(2\lambda_{Ras} - 1)n}{4\tilde{N} + (2\lambda_{Ras} - 1)n}, \\
		n_t &= -\frac{2(2\lambda_{Ras} - 1)n}{4\tilde{N} + (2\lambda_{Ras} - 1)n}.
	\end{align}
\end{subequations}

 \begin{figure}[!htpb]\centering
	\includegraphics[width=7cm]{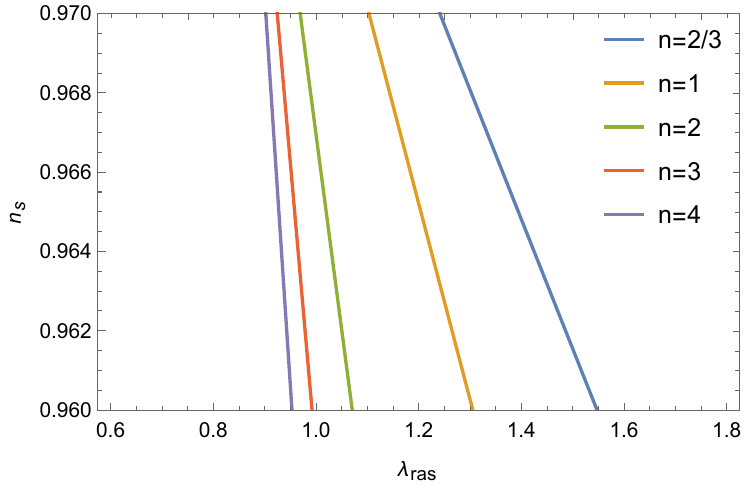}
	\includegraphics[width=7cm]{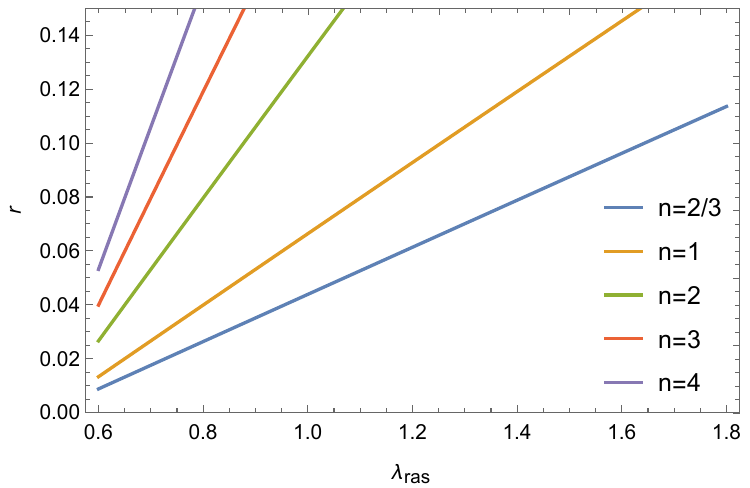}
	\includegraphics[width=7cm]{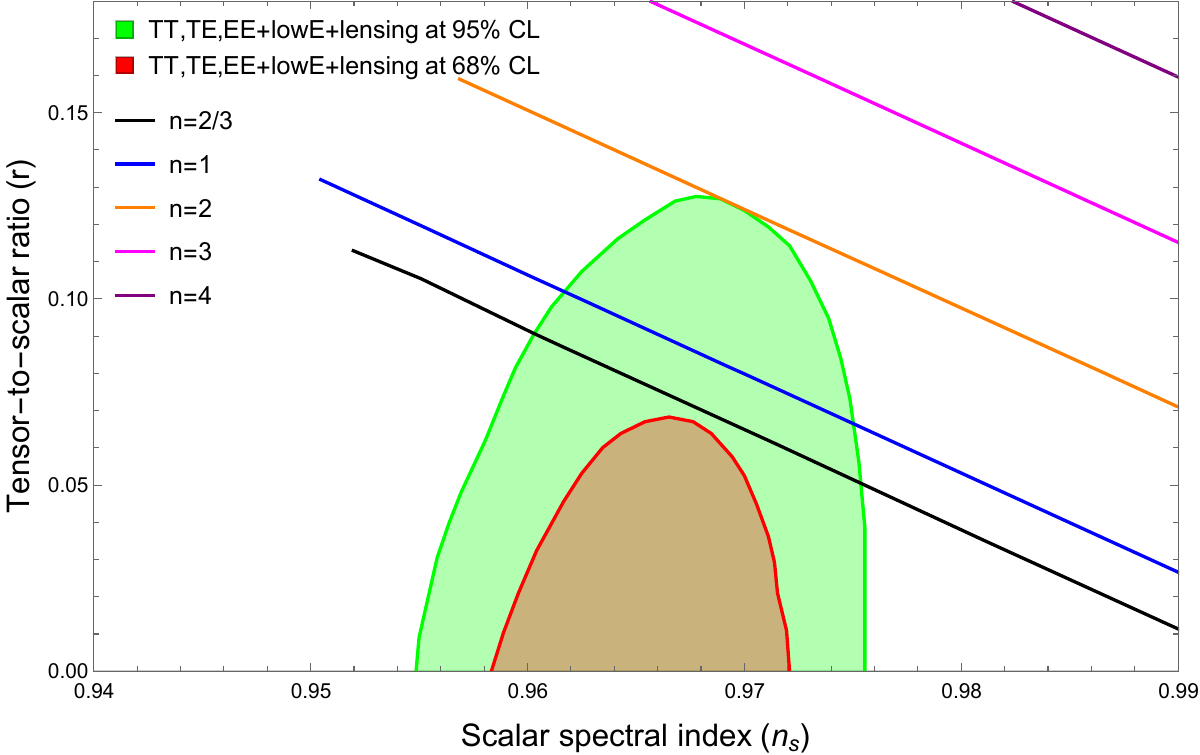}
	\caption{Evolution of the Spectral Index \(n_s\) versus \(\lambda_{Ras}\) (Left Panel), \(r\) versus \(\lambda_{Ras}\) (Right Panel), and the \(n_s - r\) Evolution (Bottom Panel) for Various \(n\) Values	
} \label{fig1}
\end{figure}

Figure \ref{fig1} clearly illustrates the results obtained for the power law potential, where five distinct cases for the parameter \(n\) (2/3, 1, 2, 3, and 4) are examined at \(N = 60\) e-folds. It is immediately evident that only the cases with \(n = 2/3\) and \(n = 1\) satisfy the Planck constraint for \(r < 0.1\), while the other cases do not fall within the bounds established by Planck+BK15 and Planck+BK18. More specifically, the \(n_s - r\) plot shows that only these two cases enter the \(2\sigma\) region of the Planck 2018 bounds, with the case \(n = 2\) just touching the boundary of this region. Furthermore, the analysis reveals that as \(\lambda_{ras}\) increases, \(n_s\) decreases, and \(r\) increases in all cases. This trend allows for a precise determination of the allowed range of \(\lambda_{ras}\): \(1.26 < \lambda_{ras} < 1.54\) for \(n = 2/3\) and \(1.11 < \lambda_{ras} < 1.26\) for \(n = 1\). These findings provide a valuable assessment of the viability of these power-law potential models under current cosmological constraints.

\subsection{Natural \& Hilltop Inflation}

\subsubsection{Natural inflation}

Let us now consider alternative potentials. In particular, we focus on Natural inflation \cite{r76}, a model in which the inflaton is identified as a pseudo-Nambu–Goldstone boson generated through both spontaneous and explicit symmetry breaking. In this scenario, the dynamics of a single field are controlled by a potential of the form
\begin{equation}
	\label{eq:Natural_Inflation_Pot}
	V(\vartheta) = \Lambda^4 \left[1+\cos\left(\frac{\vartheta}{f}\right)\right],
\end{equation}
where the mass scales \(f\) and \(\Lambda\) are introduced. It has been shown that such a potential is capable of driving inflation if \(\Lambda \sim M_{\textrm{GUT}} \sim 10^{16}\,\textrm{GeV}\) and \(f\sim M_{\textrm{Pl}}=\kappa^{-1/2}\), with \(M_{\textrm{Pl}}\) being the reduced Planck mass. Using a procedure analogous to that employed earlier, the slow-roll parameters can be expressed in terms of the number of e-folds \(N\) as follows:
\begin{align}
	\epsilon_{\textrm{V}} &= \frac{M_{\textrm{Pl}}^2}{2f^2}\,\frac{\sin^2(\vartheta/f)}{\Bigl[1+\cos(\vartheta/f)\Bigr]^2} 
	\,,\\[1mm]
	\eta_{\textrm{V}} &= - \frac{M_{\textrm{Pl}}^2}{f^2}\,\frac{\cos(\vartheta/f)}{1+\cos(\vartheta/f)}\,.
\end{align}

Here, inflation ends when the modified slow-roll parameter \(	\epsilon_{\textrm{V}} \approx 1\). 
Thus, according to the definition in \eqref{eq:f(R,T)_e-folds_definition} and \eqref{eq:Natural_Inflation_Pot}, the number of e--folds \(N\) and  is given by
\begin{equation}
	N = \frac{f^2}{M_{\textrm{Pl}}^2}\,\ln\!\left(1 + \frac{M_{\textrm{Pl}}^2}{2f^2\,\tan^2\Bigl(\frac{\vartheta}{2f}\Bigr)}\right) \,.
\end{equation}
the slow-roll parameters can be expressed in terms of the number of e-folds \(N\) as follows
\begin{align}
	\epsilon_{\textrm{V}} &=\frac{M_{\textrm{Pl}}^2}{2f^2}\,\frac{1}{e^{NM_{\textrm{Pl}}^2/f^2}-1}\,,\\[1mm]
	\eta_{\textrm{V}} &= -\frac{M_{\textrm{Pl}}^2}{2f^2}\,\frac{e^{NM_{\textrm{Pl}}^2/f^2}-2}{e^{NM_{\textrm{Pl}}^2/f^2}-1}\,.
\end{align}

Consequently, the scalar spectral index and tensor-to-scalar ratio (in the slow-roll approximation, as given by \eqref{eq:spectral_indices_GR}) become
\begin{subequations}
	\label{eq:spectral_indexes_power_law_GR}
	\begin{align}
		n_{\textrm{S}} -1 &\approx -\frac{M_{\textrm{Pl}}^2}{f^2}\,\frac{e^{NM_{\textrm{Pl}}^2/f^2}+1}{e^{NM_{\textrm{Pl}}^2/f^2}-1}\,,\\[1mm]
		n_{\textrm{T}} &\approx -\frac{M_{\textrm{Pl}}^2}{f^2}\,\frac{1}{e^{NM_{\textrm{Pl}}^2/f^2}-1}\,,\\[1mm]
		r &\approx \frac{8M_{\textrm{Pl}}^2}{f^2}\,\frac{1}{e^{NM_{\textrm{Pl}}^2/f^2}-1}\,.
	\end{align}
\end{subequations}
Moreover, one can recast \(r\) as a function of \(n_s\), yielding the representation in the \((r,n_s)\) plane:
\begin{equation}
	r(n_s) \approx 4\Bigl(1-n_s-\frac{M_{\textrm{Pl}}^2}{f^2}\Bigr).
\end{equation}

Next, we explore whether Rastall gravity induces any modifications in this model. By following a similar approach, the corrected slow-roll parameters are found to be
\begin{align}
	\tilde{\epsilon}_{\textrm{V}} &=\frac{(2\lambda_{Ras}-1)^2}{2\kappa f^2}\,\tan^2\Bigl(\frac{\vartheta}{2f}\Bigr)  
	\,,\\[1mm]
	\tilde{\eta}_{\textrm{V}} &=-\frac{(2\lambda_{Ras}-1)}{2\kappa f^2}\Bigl[1-\tan^2\Bigl(\frac{\vartheta}{2f}\Bigr)\Bigr] \,.
\end{align}

\begin{subequations}
	\begin{align}
		n_{\textrm{S}} -1 &\approx -\frac{6(2\lambda_{Ras}-1)^2}{2\kappa f^2}\,\tan^2\Bigl(\frac{\vartheta}{2f}\Bigr)
		-\frac{(2\lambda_{Ras}-1)}{\kappa f^2}\Bigl[1-\tan^2\Bigl(\frac{\vartheta}{2f}\Bigr)\Bigr]\,,\\[1mm]
		n_{\textrm{T}} &\approx -\frac{(2\lambda_{Ras}-1)^2}{\kappa f^2}\,\tan^2\Bigl(\frac{\vartheta}{2f}\Bigr) \,,\\[1mm]
		r &\approx \frac{8(2\lambda_{Ras}-1)^2}{\kappa f^2}\,\tan^2\Bigl(\frac{\vartheta}{2f}\Bigr)\,.
	\end{align}
\end{subequations}

Here, inflation ends when the modified slow-roll parameter \(\tilde{\epsilon}_{\textrm{V}} \approx 1\), implying that
\begin{equation}
	\vartheta_{\textrm{f}} =2f\,\arctan\Bigl(\sqrt{\frac{2\kappa f^2}{(2\lambda_{Ras}-1)^2}}\Bigr) \,.
\end{equation}
Thus, according to the definition in \eqref{eq:f(R,T)_e-folds_definition} and \eqref{eq:Natural_Inflation_Pot}, the number of e--folds \(\tilde{N}\) and  is given by
\begin{equation}
	\tilde{N} = \frac{2\kappa f^2}{(2\lambda_{Ras}-1)}\,\ln\left[\frac{\sin\Bigl(\frac{\vartheta_f}{2f}\Bigr)}{\sin\Bigl(\frac{\vartheta}{2f}\Bigr)}\right] \,.
\end{equation}
the slow-roll parameters can be expressed in terms of the number of e-folds \(N\) as follows
\begin{align}
	\tilde{\epsilon}_V(\tilde{N})
	&=&\frac{(2\lambda_{Ras}-1)^2}{2\kappa f^2}\,
	\frac{\Delta\,e^{-\frac{(2\lambda_{Ras}-1)}{\kappa f^2}\tilde{N}}}{1-\Delta\,e^{-\frac{(2\lambda_{Ras}-1)}{\kappa f^2}\tilde{N}}}\,,\\[1mm]
	\tilde{\eta}_V(\tilde{N})
	&=&-\frac{(2\lambda_{Ras}-1)}{2\kappa f^2}\,
	\left[1-\frac{\Delta\,e^{-\frac{(2\lambda_{Ras}-1)}{\kappa f^2}\tilde{N}}}{1-\Delta\,e^{-\frac{(2\lambda_{Ras}-1)}{\kappa f^2}\tilde{N}}}\right]\,.
\end{align}
Accordingly, the spectral indices now become
\begin{subequations}
	\begin{align}
		n_{\textrm{S}} -1 &\approx -\frac{(2\lambda_{Ras}-1)}{\kappa f^2}\,\frac{2(3\lambda_{Ras}-2)\,\Delta\,e^{-\frac{(2\lambda_{Ras}-1)}{\kappa f^2}\tilde{N}}+1-\Delta\,e^{-\frac{(2\lambda_{Ras}-1)}{\kappa f^2}\tilde{N}}}{1-\Delta\,e^{-\frac{(2\lambda_{Ras}-1)}{\kappa f^2}\tilde{N}}}\,,\\[1mm]
		n_{\textrm{T}} &\approx -\frac{(2\lambda_{Ras}-1)^2}{\kappa f^2}\,
		\frac{\Delta\,e^{-\frac{(2\lambda_{Ras}-1)}{\kappa f^2}\tilde{N}}}{1-\Delta\,e^{-\frac{(2\lambda_{Ras}-1)}{\kappa f^2}\tilde{N}}}\,,\\[1mm]
		r &\approx \frac{8(2\lambda_{Ras}-1)^2}{\kappa f^2}\,
		\frac{\Delta\,e^{-\frac{(2\lambda_{Ras}-1)}{\kappa f^2}\tilde{N}}}{1-\Delta\,e^{-\frac{(2\lambda_{Ras}-1)}{\kappa f^2}\tilde{N}}}\,.
	\end{align}
\end{subequations}
where \begin{equation}
\Delta=\frac{2\kappa f^2}{(2\lambda_{Ras}-1)^2+2\kappa f^2}\,.
\end{equation}

 \begin{figure}[!htpb]\centering
	\includegraphics[width=8cm]{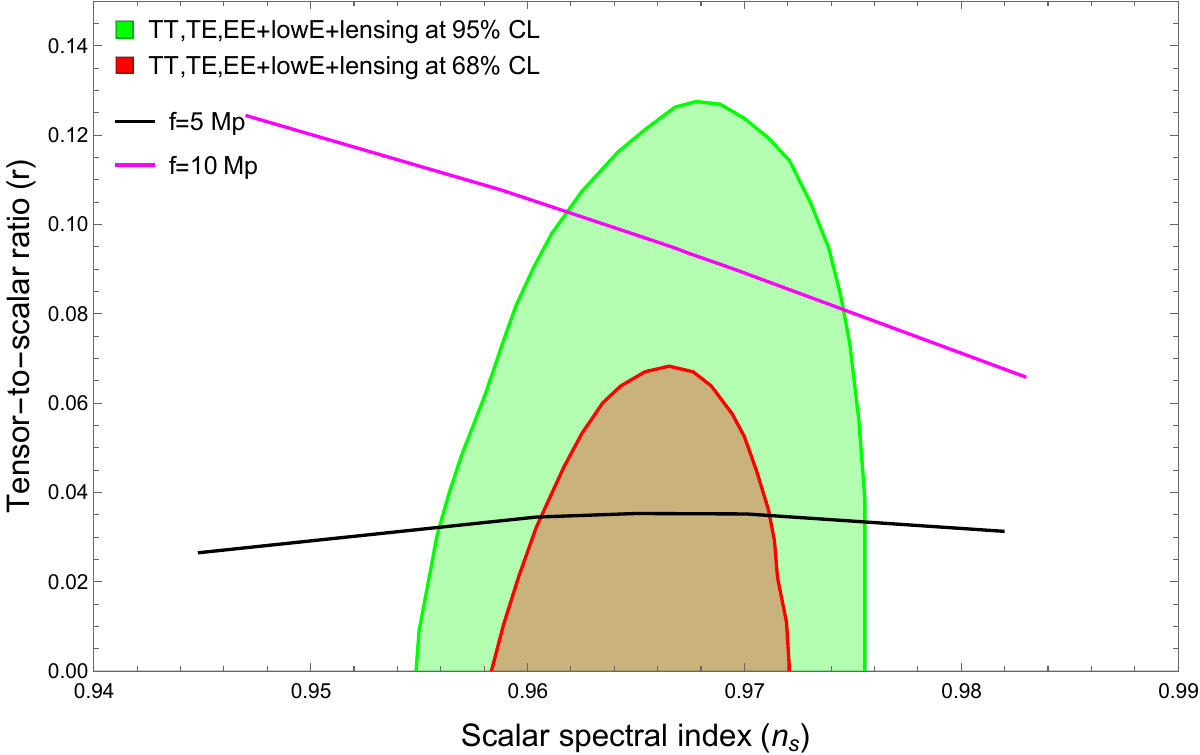}
	\caption{The \(n_s - r\) evolution for various \(f\) values	
	} \label{fig2}
\end{figure}

Figure \ref{fig2} illustrates the numerical analysis of the natural potential for $60$ e-folds by comparing two distinct mass scales. For \( f = 10M_p \), the calculated values of \( n_s \) and \( r \) satisfy only the Planck 2018 constraints, with the trajectory in the \( n_s - r \) plane entering the \( 2\sigma \) region. In this configuration, to achieve \( r < 0.1 \), the model parameter must range between 0.95 and 1.04, demonstrating that an increase in \( \lambda_{Ras} \) leads to a decrease in \( n_s \) and an increase in \( r \). In contrast, for \( f = 5M_p \), the values of \( n_s \) and \( r \) not only satisfy the Planck 2018 constraints but also adhere to the more stringent Planck + BK15 bound, positioning the curve in the \( 1\sigma \) region. Here, the model parameter range is determined to be between 0.8 and 0.9 for \( r < 0.1 \), and this range also holds for the constraint \( r < 0.056 \). Furthermore, attempts to lower the mass scale to \( f = 0.1M_p \) resulted in an insufficient number of e-folds, underlining the crucial role of the mass scale in ensuring the model's viability with respect to cosmological observations.

\subsubsection{Hilltop Potentials}

Hilltop potentials were originally introduced by Lotfi Boubekeur and David H. Lyth in 2005 \cite{r77}. In this class of models, the potential resembles a hill, with the inflaton rolling down from its peak toward the minimum, hence the term “hilltop potential”. The general form of such a potential is given by
\begin{equation}
	V(\vartheta)=\Lambda^4 \left[1-\Bigl(\frac{\vartheta}{\mu}\Bigr)^m + \dots\right],
	\label{37}
\end{equation}
where \(\Lambda\) represents the inflationary scale, \(\mu\) is the vacuum expectation value of the inflaton, and \(m\) is a positive integer. Hilltop potentials generally match the  current observational bounds at super Planckian scales. However, in some modified gravity frameworks, this scale can be lower down to the Planck scale \cite{r78}.

 Using the definitions of \(\epsilon_{\textrm{V}}\) and \(\eta_{\textrm{V}}\) provided in equations \eqref{eq:epsilon_V_GR} and \eqref{eq:eta_V_GR}, we find the slow-roll parameters to be
\begin{equation}
	\begin{aligned}
		\epsilon_{\textrm{V}} &= \frac{m^2}{2\kappa}\,\frac{\vartheta^{2(m-1)}}{\mu^{2m}\left[1-(\vartheta/\mu)^m\right]^2}\,,\\[1mm]
		\eta_{\textrm{V}} &= -\frac{m(m-1)}{\kappa}\,\frac{\vartheta^{\,m-2}}{\mu^m\left[1-(\vartheta/\mu)^m\right]}\,.
	\end{aligned}
\end{equation}

Furthermore, the spectral indices for this model are derived from \eqref{eq:spectral_indices_GR}
\begin{equation}
	\begin{aligned}
		n_s &= 1 -\frac{3m^2}{\kappa}\,\frac{\vartheta^{2(m-1)}}{\mu^{2m}\left[1-(\vartheta/\mu)^m\right]^2}
		-\frac{2m(m-1)}{\kappa}\,\frac{\vartheta^{\,m-2}}{\mu^m\left[1-(\vartheta/\mu)^m\right]}\,,\\[2mm]
		r &= \frac{8m^2}{\kappa}\,\frac{\vartheta^{2(m-1)}}{\mu^{2m}\left[1-(\vartheta/\mu)^m\right]^2}\,,\\[2mm]
		n_t &= -\frac{m^2}{\kappa}\,\frac{\vartheta^{2(m-1)}}{\mu^{2m}\left[1-(\vartheta/\mu)^m\right]^2}\,.
	\end{aligned}
\end{equation}
Next, we evaluate how Rastall Gravity modifies the slow-roll parameters and the spectral indices. From equations \eqref{epsilonv}, \eqref{etav}, and \eqref{spectralindex}, we obtain

\begin{equation}
	\begin{aligned}
		\tilde{\epsilon}_{\textrm{V}} &=
		\frac{(2\lambda_{Ras}-1)^2}{2\kappa}\,\frac{m^2\,\vartheta^{2m-2}}{\mu^{2m}} ,\\[1mm]
		\tilde{\eta}_{\textrm{V}} &=-\frac{(2\lambda_{Ras}-1)m(m-1)}{\kappa}\,\frac{\vartheta^{m-2}}{\mu^m} ,\\[2mm]
		n_s &=1-\frac{3\,(2\lambda_{Ras}-1)^2\,m^2}{\kappa}\,\frac{\vartheta^{2m-2}}{\mu^{2m}}
		-\frac{2\,(2\lambda_{Ras}-1)\,m(m-1)}{\kappa}\,\frac{\vartheta^{m-2}}{\mu^m} ,\\[2mm]
		r &=\frac{8\,(2\lambda_{Ras}-1)^2\,m^2}{\kappa}\,\frac{\vartheta^{2m-2}}{\mu^{2m}}.
	\end{aligned}
\end{equation}
The end of inflation is determined by  make using
\begin{equation}
\tilde{\epsilon}_V(\vartheta_f)=1.
\end{equation}
In this way, we can evaluate between \(\vartheta_f\) and \(\vartheta\), the number of e–folds \(\tilde{N}\).

-For \(m\neq2\):

\begin{equation}
	\tilde{N} = \frac{\kappa\,\mu^m}{m\,(2\lambda_{Ras}-1)(m-2)}\left(\vartheta^{-(m-2)}-\vartheta_f^{-(m-2)}\right)
\end{equation}

with
\begin{equation}
	\vartheta_f=\left[\frac{2\kappa\,\mu^{2m}}{(2\lambda_{Ras}-1)^2\,m^2}\right]^{\frac{1}{2m-2}}.
\end{equation}

- For \(m=2\):

\begin{equation}
	\tilde{N}=\frac{\kappa\,\mu^2}{2(2\lambda_{Ras}-1)}\,\ln\left(\frac{\sqrt{2\kappa}\,(2\lambda_{Ras}-1)\,\vartheta}{\mu^2}\right),
\end{equation}

with
\begin{equation}
	\vartheta_f=\frac{\mu^2}{\sqrt{2\kappa}\,(2\lambda_{Ras}-1)}.
\end{equation}

By substituting the explicit expressions for the slow‐roll parameters into the standard formulas, we obtain the following explicit expressions for the spectral indices in terms of the field value \( \vartheta(\tilde{N}) \) (which is implicitly a function of the number of e–folds \(\tilde{N}\)) and the parameter \(\lambda_{Ras}\)

\subsection*{Case 1: $m\neq2$}

We start from
\begin{equation}
	\vartheta = \left[\frac{m(2\lambda_{Ras}-1)(m-2)}{\kappa\,\mu^m}\,\tilde{N} + \vartheta_f^{-(m-2)}\right]^{-\frac{1}{m-2}}, \quad \text{with} \quad \vartheta_f = \left[\frac{2\kappa\,\mu^{2m}}{(2\lambda_{Ras}-1)^2\,m^2}\right]^{\frac{1}{2m-2}}.
\end{equation}
In general, for any exponent $k$ we have
\begin{equation}
	\vartheta^k = \left[\frac{m(2\lambda_{Ras}-1)(m-2)}{\kappa\,\mu^m}\,\tilde{N} + \vartheta_f^{-(m-2)}\right]^{-\frac{k}{m-2}}.
\end{equation}
In particular, setting:
\begin{itemize}
	\item $k=2m-2$ yields
	\begin{equation}
		\vartheta^{2m-2} = \left[\frac{m(2\lambda_{Ras}-1)(m-2)}{\kappa\,\mu^m}\,\tilde{N} + \vartheta_f^{-(m-2)}\right]^{-\frac{2m-2}{m-2}},
	\end{equation}
	\item and $k=m-2$ gives
	\begin{equation}
		\vartheta^{m-2} = \left[\frac{m(2\lambda_{Ras}-1)(m-2)}{\kappa\,\mu^m}\,\tilde{N} + \vartheta_f^{-(m-2)}\right]^{-1}.
	\end{equation}
\end{itemize}

Using these relations, the expressions become:

\begin{enumerate}
	\item For $\tilde{\epsilon}_V$:
	\begin{equation}
		\tilde{\epsilon}_{V}\simeq \frac{(2\lambda_{Ras}-1)^2}{2\kappa}\,\frac{m^2}{\mu^{2m}}
		\left[\frac{m(2\lambda_{Ras}-1)(m-2)}{\kappa\,\mu^m}\,\tilde{N} + \vartheta_f^{-(m-2)}\right]^{-\frac{2m-2}{m-2}}.
	\end{equation}
	\item For $\tilde{\eta}_V$:
	\begin{equation}
		\tilde{\eta}_{V}\simeq -\frac{(2\lambda_{Ras}-1)m(m-1)}{\kappa\,\mu^m}
		\left[\frac{m(2\lambda_{Ras}-1)(m-2)}{\kappa\,\mu^m}\,\tilde{N} + \vartheta_f^{-(m-2)}\right]^{-1}.
	\end{equation}
	\item For $r$:
	\begin{equation}
		r\simeq\frac{8\,(2\lambda_{Ras}-1)^2\,m^2}{\kappa\,\mu^{2m}}
		\left[\frac{m(2\lambda_{Ras}-1)(m-2)}{\kappa\,\mu^m}\,\tilde{N} + \vartheta_f^{-(m-2)}\right]^{-\frac{2m-2}{m-2}}\,.
	\end{equation}
	\item For $n_s$:
	\begin{equation}
		n_s\simeq1-\frac{3\,(2\lambda_{Ras}-1)^2\,m^2}{\kappa\,\mu^{2m}}
		\left[\frac{m(2\lambda_{Ras}-1)(m-2)}{\kappa\,\mu^m}\,\tilde{N} + \vartheta_f^{-(m-2)}\right]^{-\frac{2m-2}{m-2}}
	 -\frac{2\,(2\lambda_{Ras}-1)m(m-1)}{\kappa\,\mu^m}
		\left[\frac{m(2\lambda_{Ras}-1)(m-2)}{\kappa\,\mu^m}\,\tilde{N} + \vartheta_f^{-(m-2)}\right]^{-1}\,.
	\end{equation}
\end{enumerate}

\subsection*{Case 2: $m=2$}

For this case the relation is
\begin{equation}
	\vartheta = \vartheta_f\,\exp\!\left(\frac{2(2\lambda_{Ras}-1)}{\kappa\,\mu^2}\,\tilde{N}\right), \quad \text{with} \quad \vartheta_f=\frac{\mu^2}{\sqrt{2\kappa}\,(2\lambda_{Ras}-1)}.
\end{equation}
Since here $2m-2=2$ and $m-2=0$ (with $\vartheta^0=1$), the expressions simplify as follows:

\begin{enumerate}
	\item For $\tilde{\epsilon}_V$:
	\begin{equation}
		\tilde{\epsilon}_{V}\simeq \frac{(2\lambda_{Ras}-1)^2}{2\kappa}\,\frac{2^2}{\mu^4}\,\vartheta^2
		= \frac{2(2\lambda_{Ras}-1)^2}{\kappa}\,\frac{\vartheta^2}{\mu^4}\,,
	\end{equation}
	where
	\begin{equation}
		\vartheta^2 = \vartheta_f^2\,\exp\!\left(\frac{4(2\lambda_{Ras}-1)}{\kappa\,\mu^2}\,\tilde{N}\right).
	\end{equation}
	\item For $\tilde{\eta}_V$:
	\begin{equation}
		\tilde{\eta}_{V}\simeq -\frac{(2\lambda_{Ras}-1)\,2\,(2-1)}{\kappa\,\mu^2}
		= -\frac{2(2\lambda_{Ras}-1)}{\kappa\,\mu^2}\,.
	\end{equation}
	\item For $r$:
	\begin{equation}
		r\simeq \frac{8\,(2\lambda_{Ras}-1)^2\,2^2}{\kappa\,\mu^4}\,\vartheta^2
		= \frac{32(2\lambda_{Ras}-1)^2}{\kappa\,\mu^4}\,\vartheta_f^2\,\exp\!\left(\frac{4(2\lambda_{Ras}-1)}{\kappa\,\mu^2}\,\tilde{N}\right)\,.
	\end{equation}
	\item For $n_s$:
	\begin{equation}
		\begin{aligned}
			n_s\simeq 1 &-\frac{3\,(2\lambda_{Ras}-1)^2\,2^2}{\kappa\,\mu^4}\,\vartheta^2
			- \frac{2\,(2\lambda_{Ras}-1)\,2\,(2-1)}{\kappa\,\mu^2}\\
			&= 1 - \frac{12(2\lambda_{Ras}-1)^2}{\kappa\,\mu^4}\,\vartheta^2
			- \frac{4(2\lambda_{Ras}-1)}{\kappa\,\mu^2}\\
			&= 1 - \frac{12(2\lambda_{Ras}-1)^2}{\kappa\,\mu^4}\,\vartheta_f^2\,\exp\!\left(\frac{4(2\lambda_{Ras}-1)}{\kappa\,\mu^2}\,\tilde{N}\right)
			- \frac{4(2\lambda_{Ras}-1)}{\kappa\,\mu^2}\,.
		\end{aligned}
	\end{equation}
	Hence,
	\begin{equation}
		n_s\simeq 1 - \frac{12(2\lambda_{Ras}-1)^2}{\kappa\,\mu^4}\,\vartheta_f^2\,\exp\!\left(\frac{4(2\lambda_{Ras}-1)}{\kappa\,\mu^2}\,\tilde{N}\right)
		- \frac{4(2\lambda_{Ras}-1)}{\kappa\,\mu^2}\,.
	\end{equation}
\end{enumerate}

\begin{figure}[!htpb]\centering
        \includegraphics[width=8cm]{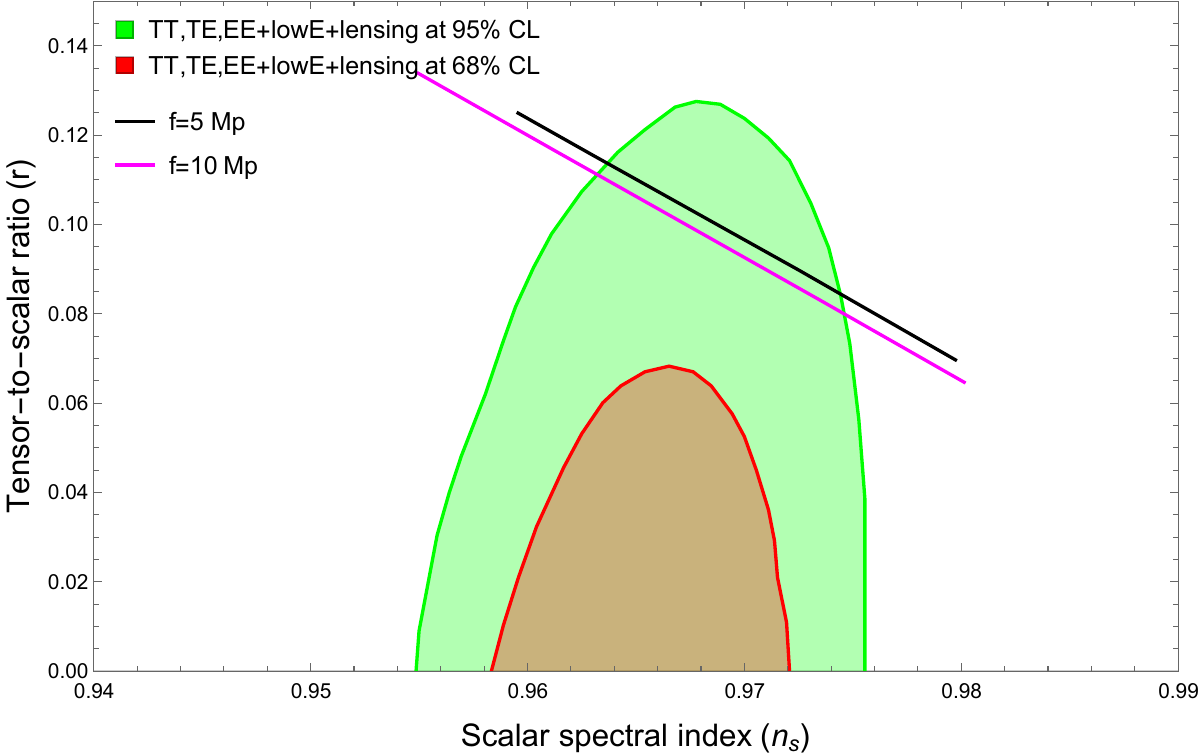}
	\includegraphics[width=8cm]{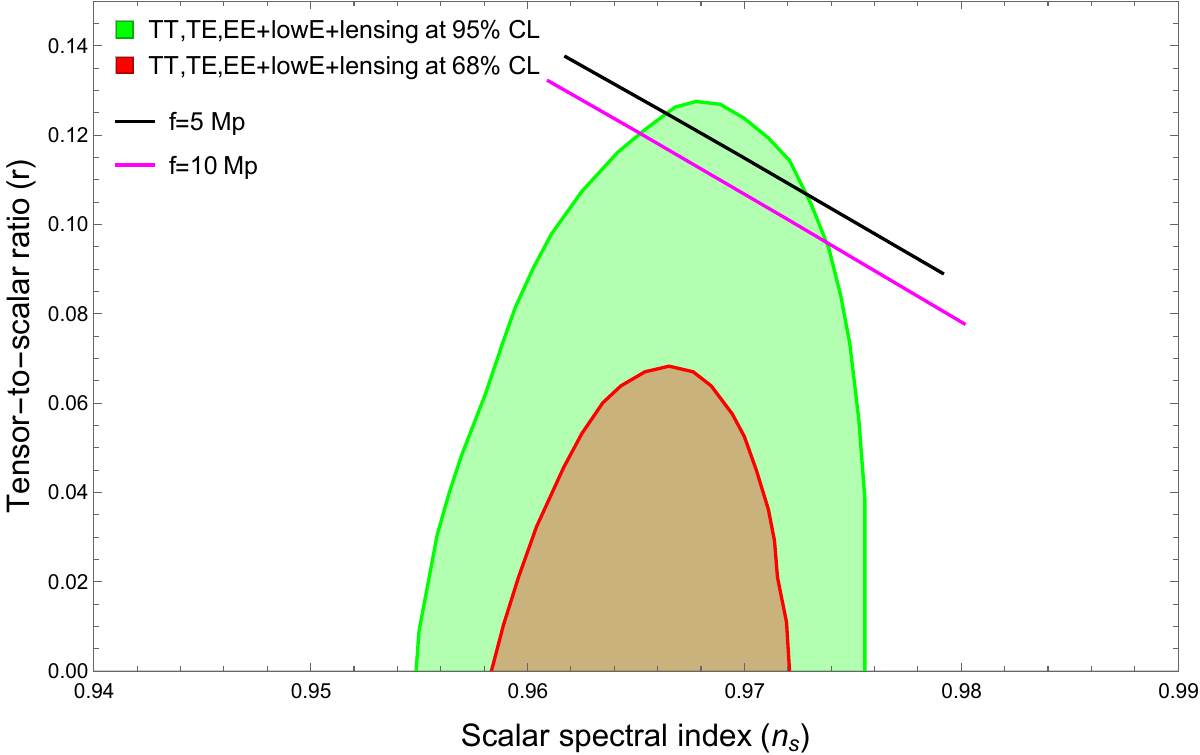}
	\includegraphics[width=8cm]{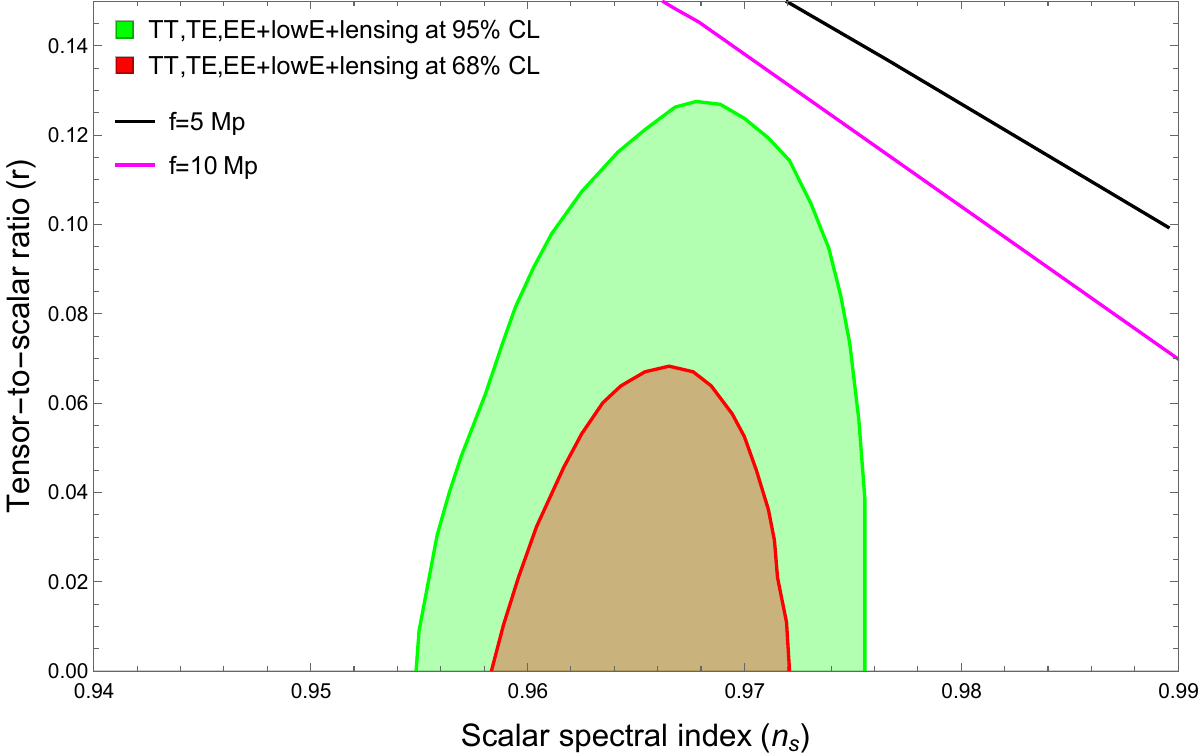}
	\includegraphics[width=8cm]{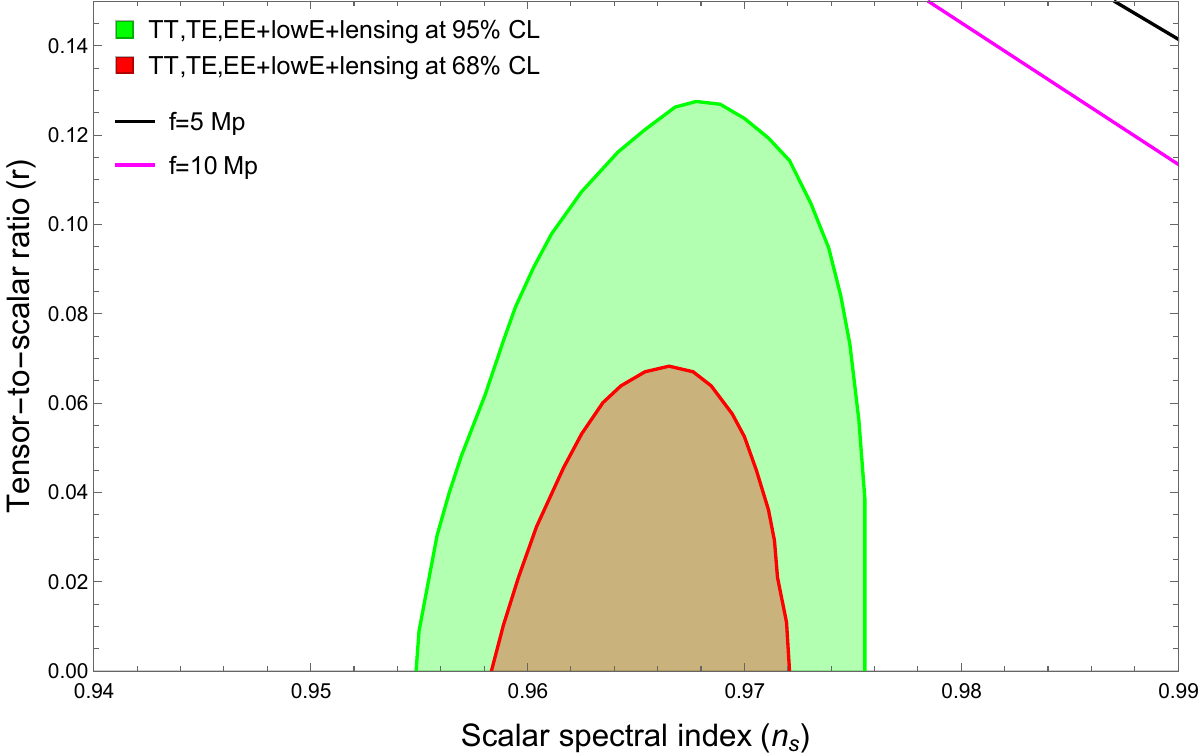}
	
	\caption{The \(n_s - r\) evolution for various cases of \(m\). The upper left one is for $m=3/2$, the upper right one is for $m=2$, the lower left one is for $m=3$, and the lower right one is for $m=4$.
	} \label{fig3}
\end{figure}

Figure \ref{fig3} illustrates the detailed results obtained from the analysis of the Hilltop potential for four distinct cases corresponding to \(m = 3/2\), \(2\), \(3\), and \(4\). For a fixed number of e-folds, \(\tilde{N} = 60\), both the scalar spectral index and the tensor-to-scalar ratio are numerically evaluated, showing a clear dependency on the parameter \(m\). The model is tested at the two energy scales, \(\mu = 10\,M_p\) and \(\mu = 5\,M_p\), which highlights the sensitivity of the predictions to the mass scale. This comparative approach underlines the significance of the Hilltop potential configuration in the cosmic inflation scenario, demonstrating how even subtle variations in the potential parameters can have a substantial impact on the overall dynamics of the universe's expansion.

\begin{figure}[!htpb]\centering
	\includegraphics[width=8cm]{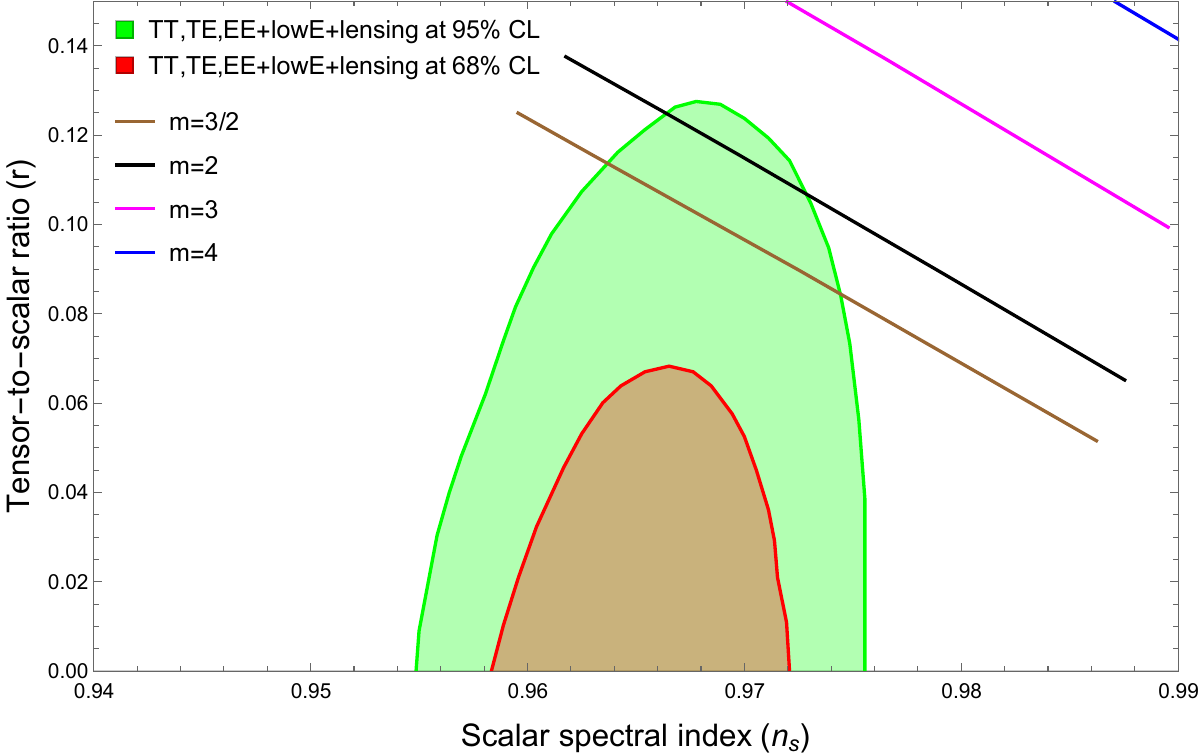}
	\caption{The \(n_s - r\) plot for different cases at \(f=5M_p\) scale
	} \label{fig4}
\end{figure}

Figure \ref{fig4} illustrates the Hilltop potential results for the case when \(\mu = 10M_p\). For \(m = \frac{3}{2}\), a viable parameter range of \(1.07 < \lambda_{Ras} < 1.11\) is obtained under condition \(r < 0.1\), with the corresponding \(n_s\)–\(r\) trajectory entering the \(2\sigma\) region for Planck 2018. However, it fails to meet the combined constraints from Planck + BK15 and Planck + BK18. In the scenario where \(m = 2\), although the \(n_s\)–\(r\) plot again touches the \(2\sigma\) region for Planck 2018, the predicted values of \(r\) exceed acceptable limits for the observational range \(n_s = 0.9649 \pm 0.0042\), resulting in no acceptable parameter range. Furthermore, for both \(m = 3\) and \(m = 4\), the \(n_s\)–\(r\) curves do not enter the \(2\sigma\) region at all, meaning that no value of \(\lambda_{Ras}\) produces results consistent with observations. This analysis highlights the critical dependence of the model’s viability on the exponent \(m\) and the Rastall parameter, with only the \(m = \frac{3}{2}\) case showing marginal compatibility under limited conditions.

\begin{figure}[!htpb]\centering
	\includegraphics[width=8cm]{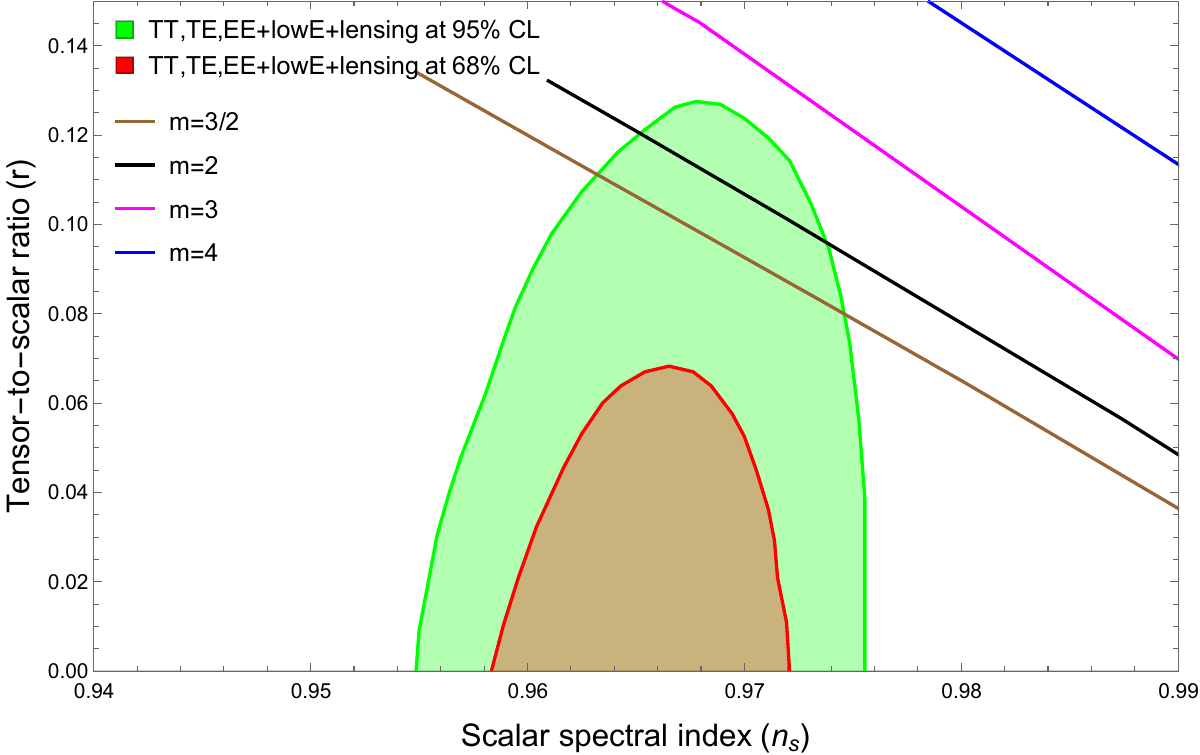}
	\caption{The \(n_s - r\) plot for different cases at \(f=10M_p\) scale
	} \label{fig5}
\end{figure}

Figure \ref{fig5} illustrates the results obtained from the hilltop potential with \( \mu = 5M_p \). For \( m = \frac{3}{2} \), the analysis reveals that the parameter range \( 1.058 < \lambda_{Ras} < 1.065 \) is compatible with \( r < 0.1 \), yet the resulting \( n_s - r \) plot reaches the \( 2\sigma \) region as determined by the Planck 2018 data, failing to comply with the tighter constraints imposed by Planck + BK15 and Planck + BK18. In the case of \( m = 2 \), although the \( n_s - r \) curve enters the \( 2\sigma \) region for Planck 2018, the corresponding values of \( r \) are relatively high for the \( n_s = 0.9649 \pm 0.0042 \) range, thereby ruling out any acceptable parameter space. Moreover, for \( m = 3 \) and \( m = 4 \), the \( n_s - r \) plots do not intersect the \( 2\sigma \) region at all, indicating that no viable values of \( \lambda_{Ras} \) can be found to produce the desired results for these cases.
\section{Conclusion}\label{sec:discussion}

In summary, our study comprehensively examined the interplay between modified gravity in the Rastall framework and slow-roll inflation through several potential models, including the power law, natural, and Hilltop potentials. For the power law potential, only the cases with \( n = \frac{2}{3} \) and \( n = 1 \) satisfy the Planck constraint \( r < 0.1 \) and fall within the \(2\sigma\) region, with precise ranges of \(\lambda_{ras}\) established for each. The analysis of the natural potential further demonstrated that the viability of the model is strongly dependent on the mass scale: while \( f = 10\,M_p \) only marginally meets the Planck 2018 criteria, \( f = 5\,M_p \) robustly satisfies both the Planck 2018, and more stringent Planck + BK15 constraints; lower mass scales are found to be inadequate due to insufficient e-folds.

 Hilltop potential analysis performed at two different energy scales (\(\mu = 10\,M_p\) and \(\mu = 5\,M_p\)), revealed that the parameter \(m\) critically influences the inflationary predictions. At \(\mu = 10\,M_p\), only the case \(m = \frac{3}{2}\) shows a narrow window of viability, while the cases \(m = 2\), \(m = 3\), and \(m = 4\) fail to meet observational bounds. A similar trend is observed at \(\mu = 5\,M_p\), reinforcing the sensitivity of the model's predictions to both the choice of \(m\) and the energy scale.

Overall, these findings highlight the delicate balance between theoretical parameters and observational constraints in inflationary cosmology. They also emphasize the potential of modified gravity theories to serve as promising alternatives in explaining the early universe dynamics. Future research should aim to further refine these parameter spaces and investigate additional observational signatures to strengthen the viability of these models.

\end{document}